# Automated segmentation of 3-D body composition on computed tomography

**Lucy Pu[a], Syed F. Ashraf[a], Naciye S Gezer[b], Iclal Ocak[b], Rajeev Dhupar[a,b*]**

[a]Department of Cardiothoracic Surgery, University of Pittsburgh School of Medicine, Pittsburgh, PA, 15213, USA

[b]Department of Radiology, University of Pittsburgh School of Medicine, Pittsburgh, PA, 15213, USA

[c]Surgical Services Division, VA Pittsburgh Healthcare System, Pittsburgh, PA, 15230, USA

**\*Corresponding author and guarantor of the entire manuscript:**

Rajeev Dhupar, MD, MBA

Assistant Professor of Cardiothoracic Surgery

University of Pittsburgh School of Medicine

Chief of Thoracic Surgery, VAPHS

Shadyside Medical Building, Suite 715

5200 Centre Avenue, Pittsburgh, PA 15232

dhuparr2@upmc.edu; Rajeev.Dhupar@va.gov

Contact Phone: (412) 623-2025

**Abstract**

**Purpose:** To develop and validate a computer tool for automatic and simultaneous segmentation of body composition depicted on computed tomography (CT) scans for the following tissues: visceral adipose (VAT), subcutaneous adipose (SAT), intermuscular adipose (IMAT), skeletal muscle (SM), and bone.

**Approach:** A cohort of 100 CT scans acquired from The Cancer Imaging Archive (TCIA) was used - 50 whole-body positron emission tomography (PET)-CTs, 25 chest, and 25 abdominal. Five different body compositions were manually annotated (VAT, SAT, IMAT, SM, and bone). A training-while-annotating strategy was used for efficiency. The UNet model was trained using the already annotated cases. Then, this model was used to enable semi-automatic annotation for the remaining cases. The 10-fold cross-validation method was used to develop and validate the performance of several convolutional neural networks (CNNs), including UNet, Recurrent Residual UNet (R2Unet), and UNet++. A 3-D patch sampling operation was used when training the CNN models. The separately trained CNN models were tested to see if they could achieve a better performance than segmenting them jointly. Paired-samples t-test was used to test for statistical significance.

**Results:** Among the three CNN models, UNet demonstrated the best overall performance in jointly segmenting the five body compositions with a Dice coefficient of $0.840\pm0.091$, $0.908\pm0.067$, $0.603\pm0.084$, $0.889\pm0.027$, and $0.884\pm0.031$, and a Jaccard index of $0.734\pm0.119$, $0.837\pm0.096$, $0.437\pm0.082$, $0.800\pm0.042$, $0.793\pm0.049$, respectively for VAT, SAT, IMAT, SM, and bone.

**Conclusion:** There were no significant differences among the CNN models in segmenting body composition, but jointly segmenting body compositions achieved a better performance than segmenting them separately.

*Rajeev Dhupar, E-mail: dhuparr2@upmc.edu; Rajeev.Dhupar@va.gov

# 1.     Introduction

Body composition refers to the radiographic "building blocks" of the human body and includes tissues such as fat, muscle, bone, and other organs. An individual's body composition, in part, can be considered to be a reflection of her/his long-term habits and lifestyle (e.g., physical exercise and diet). It is not surprising that many diseases, such as cardiovascular disease[1], diabetes[2], cancers[3], and osteoporosis[4], are closely associated with body composition. Numerous studies have highlighted how body composition can infer long- and short-term health[5-9]. Hence, accurate quantification of body composition may have significant implications for the practice of medicine. Current widely used methods for measuring or inferring body composition include body mass index, waist circumference, bioimpedance, and dual-energy x-ray absorptiometry. Compared to these traditional approaches, three-dimensional (3-D) imaging modalities (e.g., computed tomography (CT) and magnetic resonance imaging (MRI)) provide an opportunity to compute a more detailed assessment of body composition[10,11]. It was estimated that 85 million CT scans were performed in 2011 in the U.S[12]. An efficient way to automatically analyze body compositions depicted on CT images would provide a novel way to facilitate an additional and timely assessment of a patient's health status without incurring additional costs.

Although CT imaging allows volumetric quantification of body compositions, the majority of current methods quantify body composition based on the cross-sectional area of only a single image slice. This is because it is extremely time-consuming to manually segment all of the different compartments depicted on CT images. The image slice at the third cervical (C3) vertebra, third or fourth lumbar (L3 or L4) vertebra[13-16], and fourth thoracic vertebra (T4)[17] were often used as the indices of the whole-body compositions. Multiple studies have reported correlations of single image slice composition and whole body composition, however it is unclear if these can act as a consistent replacement[17-19]. Correspondingly, efforts have been made to develop automated algorithms to segment the body tissues at single locations (L3, L4, T4), however they have not been able to consistently provide results that are applicable to clinical medical practice[20-29].

In this study, we describe a convolutional neural network (CNN)-based solution to automatically segment five different body compositions depicted on any CT scan regardless of the body regions, including visceral adipose tissue (VAT), subcutaneous adipose tissue (SAT), intermuscular adipose tissue (IMAT), skeletal muscle (SM), and bone. To develop the algorithms, we collected a dataset consisting of 50 whole-body CT scans, 25 chest CT scans, and 25 abdominal CT scans. The above-mentioned five body tissues were manually annotated by an image analyst with the help of our in-house software and verified by a thoracic radiologist. During the annotation, a training-while-annotating strategy was used to improve the efficiency of the annotations. Several CNN models were trained based on 3-D patches of the CT images. A uniform sampling strategy was used to facilitate patch-based machine learning. The 10-fold cross-validation method was used to validate the performance of the trained CNN models.

# 2.     Methods and Materials

*2.1 Datasets*

We created a cohort consisting of 100 CT scans acquired on different subjects from The Cancer Imaging Archive (TCIA)[30]. Among these CT scans, 50 were the whole-body CT scans from the PET-CT examinations, 25 were chest CT examinations, and 25 were abdominal CT examinations. TCIA is a public data repository hosting a large archive of medical images of cancer as the result of a number of studies funded by the National Institute of Cancer (NCI). For the sake of data diversity, we selected these CT scans from different studies or collections (Table 1) in the TCIA archive. All these studies were performed at different medical institutes, and the CT scans were acquired using various protocols (e.g., dosage, field-of-view, slice thickness, and scanners).

**Table 1.** Summary of the cases in our cohort obtained from TCIA [30]

|  | NSCLC-Radiogenomics [31] | ACRIN-NSCLC-FDG-PET [32] | NLST [33] | C4KC-KiTS [34] |
|---|---|---|---|---|
| **Subjects (100)** | 26 | 24 | 25 | 25 |
| Men (57) | 9 | 16 | 17 | 15 |
| Female (43) | 17 | 8 | 8 | 10 |
| **Age, mean** | 65.9±11.6 | 64.4±10.6 | 62.7±4.7 | 57.2±12.4 |
| **Disease status** | Lung Cancer | Lung Cancer | miscellaneous | Kidney Cancer |
| **CT Protocol** |  |  |  |  |
| CT scan range | Whole-body | Whole-body | Lung | Abdomen |
| CT slice thickness | 3.75-5 mm | 2.5-5 mm | 2-2.5 mm | 1-5 mm |
| CT slices per scan | 205-299 | 170-453 | 100-196 | 61-417 |

An image analyst was trained by an experienced radiologist (>15 years' experience) to manually annotate five different body compositions depicted on these CT scans, including VAT, SAT, IMAT, SM, and bones, with the help of our in-house software. This in-house software supports various image operations, such as thresholding, interpolation, flooding, noise filter, morphological operators, and overlay editing (e.g., erase and painting). To improve the annotation efficiency, we developed a training-while-annotation strategy (Fig. 1). We first applied the thresholding operations to each image slice to roughly identify the body composition and then refine the boundary using other operations (e.g., overlay painting or erasing). After ten CT scans were initially annotated, the radiologist was asked to review the results. Then, we started to train a CNN model (i.e., UNet) for automated segmentation based on the annotated cases. Next, the trained UNet model was used to automatically segment the body compositions. Thereafter, the image analyst worked on the computerized results and used the in-house software to correct the improper annotations by the computer. We repeated this procedure for every ten cases until the 100 cases were fully annotated. For efficiency assessment, we uniformly divided the cases in the cohort into 10 batches with a uniform distribution of whole-body and chest/abdominal CT scans. For each batch, we recorded the average time spent on the annotations.

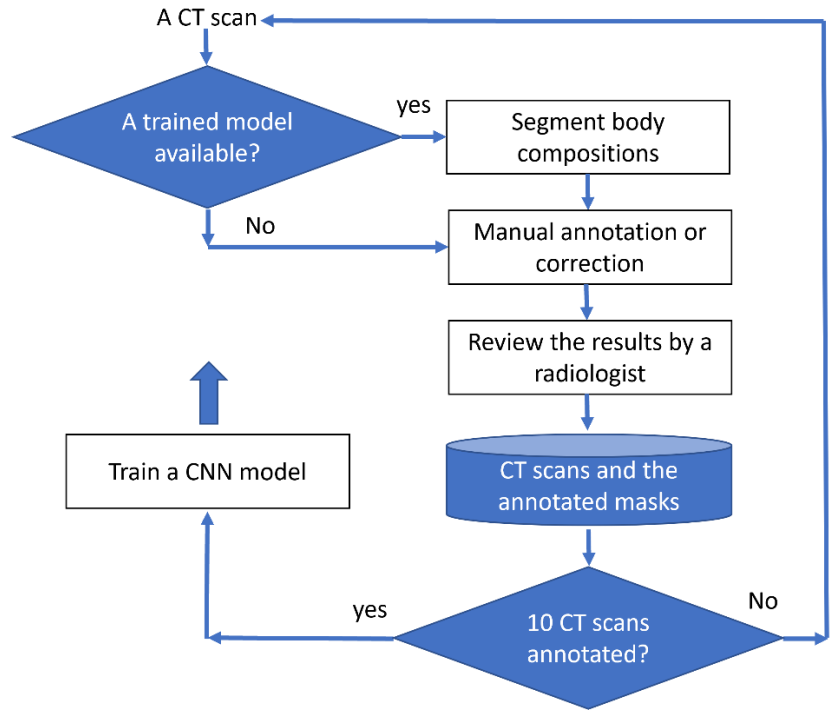

**Figure 1**. The training-while-annotating strategy for efficient annotation.

*2.2 Segmentation of body compositions depicted on CT images*

To segment the body tissues on CT images, we implemented and trained three CNN models, including UNet[35], residual recursive UNet (R2Unet)[36], and UNet++[37]. The implementation procedure was illustrated by the flowchart in Fig. 2.

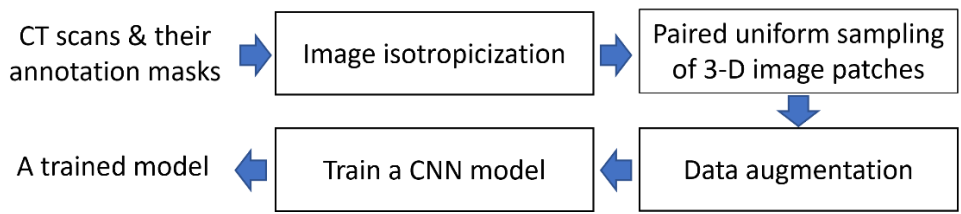

**Figure 2**. Our algorithm flowchart

First, given a CT scan and its annotation mask, we converted both of them into an isotropic form with a resolution of 1×1×1 mm$^3$.

Second, we sampled paired 3-D CT images and their corresponding mask patches from the isotropic CT scans and their annotated masks in a uniform manner. Each paired image and mask patch were located at the same location. To enable "balanced" learning, we paired the sampling of the foreground regions (i.e., body tissues of interest) and the background regions equally. In many situations, a region of interest (e.g., VAT) is often much smaller than its background. A simple random sampling may make the computer overlearn the background and thus result in a biased model. To efficiently implement this procedure, we subdivided and padded the mask into a 3-D uniform grid. Each grid had a dimension of 60×60×60 mm$^3$. Then, two grids were randomly selected. One had annotation, namely the region of interest. The other one had no or relatively few annotations. Next, we randomly selected two voxels from the selected two grids, respectively. One voxel was located in the region of interest, and the other in the background. Thereafter, we extracted the paired cubic patches centered at the two voxels from both the mask and the associated CT scan. The paired volumetric patches had a uniform dimension of 80×80×80 mm$^3$. In this study, we have five regions of interest (i.e., five body tissues). As a result, we sampled five paired cubic patches.

Third, various data augmentation operations, such as scale, rotation, translation, crop, density shift, noise addition, and Gaussian smoothing, were randomly combined and applied to the above cubic patches on the fly. The objective is to improve data diversity and thus the robustness of the trained models.

Fourth, the paired image and mask patches were fed into a 3-D CNN model. We implemented three classical CNN models, including UNet[35], residual recursive UNet (R2Unet)[36], and UNet++[37]. To train a model that can simultaneously segment the five different body tissues, we need to feed all the five paired 3-D image and mask patches into a CNN model at the same time. In these models, we used the Dice coefficient as the loss function, Adam as the optimizer, and Softmax as the activation function. The initial learning rate was set at 0.0001. If the validation performance did not increase in three epochs, the learning rate was decayed by a factor of 0.5. The training would stop if the Dice coefficient of the validation procedure was not improved for a continuous 10 epochs. Since these models have different complexity and the graphical processing unit (GPU) has limited memory (12 GB), we adjusted the batch size separately when training these models with the same dimension of 3-D patches and other same parameters (e.g., learning rate and optimizer). The batch sizes for UNet, R2Unet, and UNet++ were set at 4, 2, and 4, respectively.

*2.3 Performance validation*

We validated the performance of the CNN models using the 10-fold cross-validation method. The cohort was split into ten folds. Eight folds were used for training, one fold for internal validation, and the remaining one for an independent test. Each fold was visited in the training and independent testing. The average Dice coefficient and Jaccard index were used as the performance metrics. For simultaneous segmentation of the five body compositions, we computed the Dice coefficient for each body tissue. We used the paired sample t-test to statistically assess the performance differences among the involved CNN models. A p-value less than 0.05 was considered statistically significant. To test whether the CNN models would perform better in segmenting the body tissue separately than in segmenting all five tissues simultaneously, we also trained the CNN models separately on each body tissue. Notably, the segmentation performance was quantitatively assessed based on the original CT images and segmentation masks instead of the isotropic images and masks.

## 3. Results

As shown in Table 2, among the three CNN models, despite their relatively small differences, UNet demonstrated the best overall performance in jointly segmenting the five body tissues with a Dice coefficient of 0.840±0.091, 0.908±0.067, 0.603±0.084, 0.889±0.027, and 0.884±0.031, and a Jaccard index of 0.734±0.119, 0.837±0.096, 0.437±0.082, 0.800±0.042, 0.793±0.049, respectively for VAT, SAT, IMAT, SM, and bone. UNet++ demonstrated the poorest overall performance in segmenting these body tissues with a Dice coefficient of 0.826±0.104, 0.901±0.075, 0.574±0.092, 0.886±0.028, and 0.870±0.033 and a Jaccard index of 0.716±0.134, 0.827±0.104, 0.408±0.088, 0.796±0.045, and 0.772±0.051, respectively for VAT, SAT, IMAT, SM, and bone. There were no significant performance differences between R2Unet and UNet++ except for segmenting IMAT ($p<0.05$). Among the five body tissues, all these CNN models demonstrated the best performance in segmenting SAT and the poorest performance in segmenting IMAT

**Table 2**. Summary of the performance of three CNN models in segmenting five different body compositions simultaneously in our cohort using the 10-fold cross-validation method.

| Method | Metrics | VAT | SAT | IMAT | SM | Bone |
|---|---|---|---|---|---|---|
| UNet | Dice coefficient | 0.840±0.091 | 0.908±0.067 | 0.603±0.084 | 0.889±0.027 | 0.884±0.031 |
| | Jaccard index | 0.734±0.119 | 0.837±0.096 | 0.437±0.082 | 0.800±0.042 | 0.793±0.049 |
| R2Unet | Dice coefficient | 0.828±0.090 | 0.903±0.072 | 0.611±0.081 | 0.874±0.030 | 0.883±0.034 |
| | Jaccard index | 0.715±0.115 | 0.829±0.101 | 0.444±0.081 | 0.778±0.046 | 0.793±0.053 |
| UNet++ | Dice coefficient | 0.826±0.104 | 0.901±0.075 | 0.574±0.092 | 0.886±0.028 | 0.870±0.033 |
| | Jaccard index | 0.716±0.134 | 0.827±0.104 | 0.408±0.088 | 0.796±0.045 | 0.772±0.051 |

VAT- visceral adipose tissue; SAT-subcutaneous adipose tissue; IMAT-intermuscular adipose tissue; SM-skeletal muscle (SM)

As compared with the CNN models trained for segmenting the body tissues simultaneously, the CNN models trained for segmenting the body tissues separately demonstrated worse performance ($p<0.01$) for all tissue types. UNet still had the best segmentation performance with a Dice coefficient of 0.819±0.099, 0.896±0.082, 0.590±0.089, 0.871±0.030, and 0.870±0.036 and a Jaccard index of 0.705±0.126, 0.819±0.109, 0.424±0.086, 0.772±0.046, 0.771±0.055,

respectively for VAT, SAT, IMAT, SM, and bone. Again, R2Unet and UNet++ had similar overall segmentation performance.

Table 3. Summary of the performance of three CNN models in simultaneously segmenting five different body tissues separately in our cohort using the 10-fold cross-validation method.

| Method | Metrics | VAT | SAT | IMAT | SM | Bone |
|---|---|---|---|---|---|---|
| UNet | Dice coefficient | 0.819±0.099 | 0.896±0.082 | 0.590±0.089 | 0.871±0.030 | 0.870±0.036 |
| | Jaccard index | 0.705±0.126 | 0.819±0.109 | 0.424±0.086 | 0.772±0.046 | 0.771±0.055 |
| R2Unet | Dice coefficient | 0.744±0.154 | 0.863±0.117 | 0.433±0.109 | 0.857±0.033 | 0.808±0.045 |
| | Jaccard index | 0.613±0.175 | 0.773±0.140 | 0.282±0.089 | 0.750±0.050 | 0.680±0.061 |
| UNet++ | Dice coefficient | 0.749±0.142 | 0.856±0.123 | 0.469±0.103 | 0.838±0.035 | 0.803±0.052 |
| | Jaccard index | 0.616±0.162 | 0.763±0.143 | 0.312±0.088 | 0.723±0.051 | 0.677±0.069 |

VAT- visceral adipose tissue; SAT-subcutaneous adipose tissue; IMAT-intermuscular adipose tissue; SM-skeletal muscle (SM)

To visually demonstrate the performance of the CNN model in segmenting the five body tissues, two different examples were shown in Figs. 3 and 4. The example in Fig. 3 had a relatively high volume of SAT, while the example in Fig. 4 had a very limited amount of SAT.

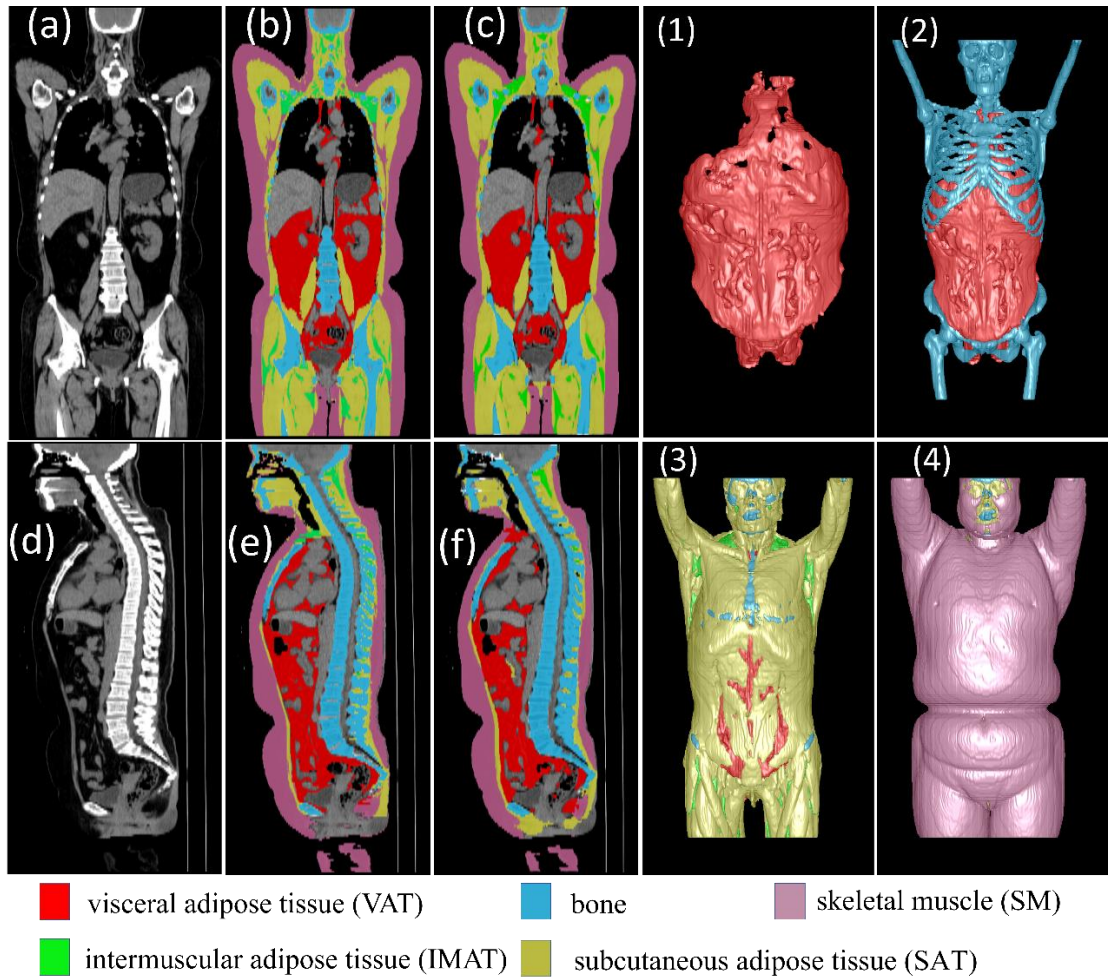

**Figure 3**. The computerized results of the identified body compositions on a subject with a relatively high volume of fat. (a)-(c) and (d)-(f) showed the coronal and sagittal views of the whole-body CT scans with the body compositions in different overlays. (a) and (d) showed the original CT images, (b) and (e) showed the manual annotations of the body compositions, and (c) and (f) showed the computerized segmentations of the body compositions. (1)-(4) showed the 3-D visualization of the body compositions.

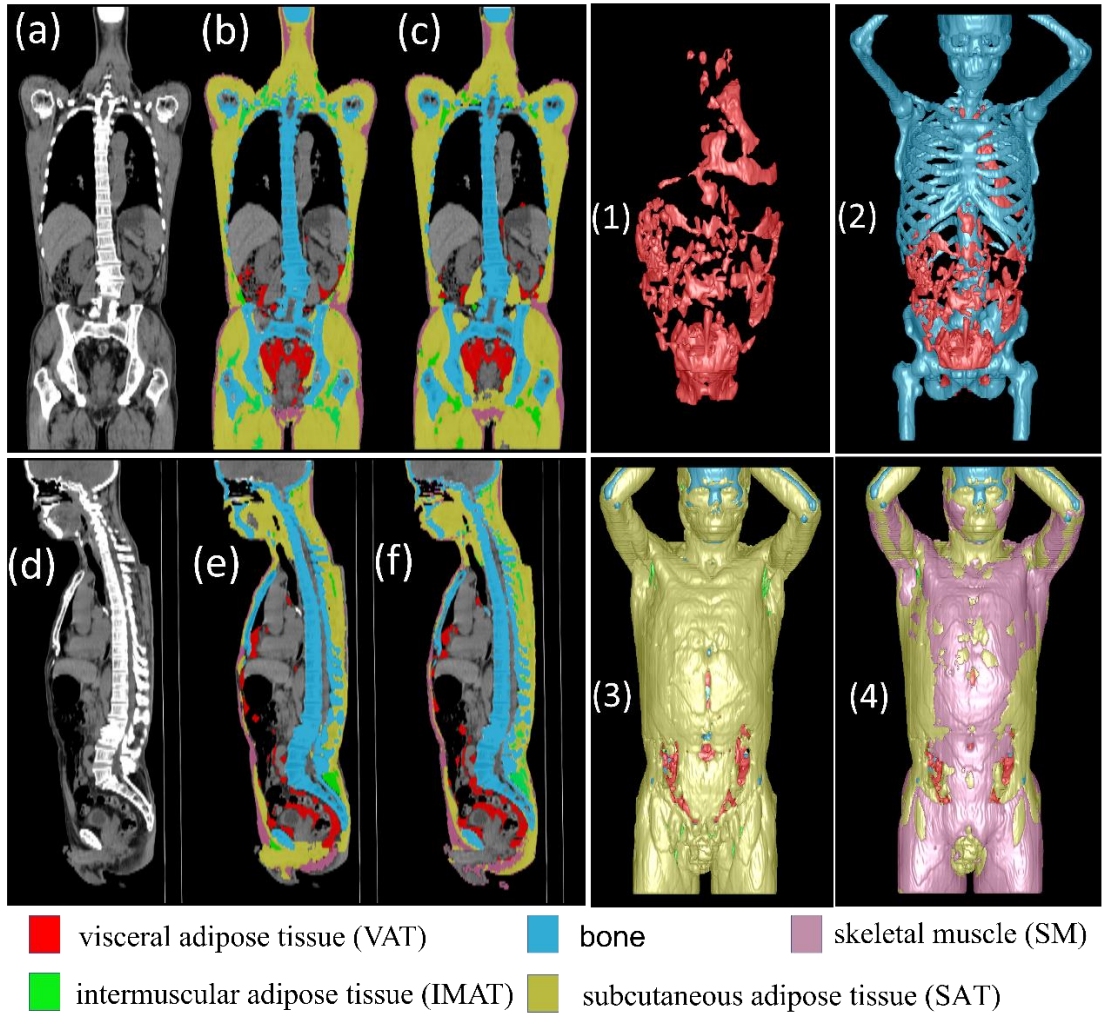

**Figure 4**. The computerized results of the identified body compositions on a subject with very little fat. (a)-(c) and (d)-(f) showed the coronal and sagittal views of the whole-body CT scans with the body compositions in different overlays. (a) and (d) showed the original CT images, (b) and (e) showed the manual annotations of the body compositions, and (c) and (f) showed the computerized segmentations of the body compositions. (1)-(4) showed the 3-D visualization of the body compositions.

Fig. 5 shows the annotation time we spent on the ten image batches when using the training-while-annotating strategy. It took about 7.5 days to annotate the first batch of CT scans. Based on the trend of the curve in Fig. 5, the efficiency of the annotation was stable after 70 cases. Thereafter, it took only about half a day to review and correct 10 CT scans. Overall, it took the image analyst 27.5 days to annotate the 100 CT scans in our cohort.

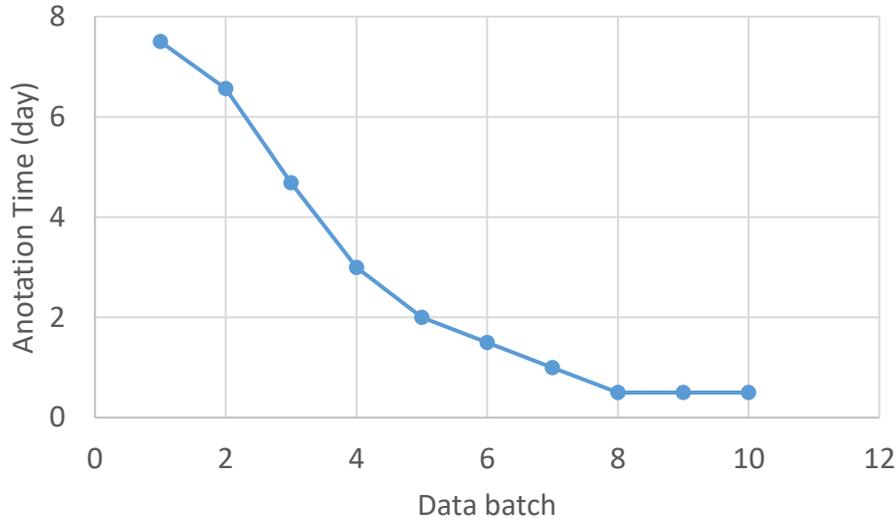

**Figure 5**. The summary of the time (days) spent on data annotations. Based on our computer configuration (GPU: NVIDIA Titan Xp - 12 GB and CPU: Intel Xeon – 3.60 GHz), it took about 2–4 minutes for the CNN models to automatically segment the body compositions depicted on a CT scan, depending on the field of view and dimension.

## 4. Discussion

We developed and validated several CNN models for automatically segmenting five different body tissues depicted on CT images. This study has several merits. First, the developed CNN models can identify five different body compositions regardless of the body region. This characteristic is primarily attributed to the diverse dataset used for machine learning, which includes CT scans acquired using a variety of protocols and covering the whole body, chest, and abdominal regions. To our knowledge, no available study has such a characteristic. In contrast, most current available algorithms can only segment 1-3 types of body composition and most focus on abdominal CT scans. Second, we developed a training-while-annotating strategy to accelerate the annotation. Based on our experience, it took approximately one day to annotate the body compositions on a CT scan. With this strategy, the annotation efficiency was improved at least four times (i.e., 2 hours per CT scan) for the cohort. Third, we developed and validated three classical CNN models. We in particular clarified whether the CNN models could perform better in segmenting the body tissues separately than in segmenting the body tissues simultaneously. These CNN models were trained and validated on a cohort formed by 100 diverse CT scans acquired on different subjects using different protocols. Finally, to improve the reliability of the trained CNN models, we used a grid-based sampling method to ensure a uniform sampling of paired 3-D image patches. The objective is to guarantee "fair" or "balanced" learning of the background and the regions of interest.

The CNN-based deep learning method is data-hungry. It needs large, carefully labeled cases to train the CNN models. Although there are investigative efforts dedicated to developing machine learning methods that can learn with limited samples, very little progress has been made. To achieve robust performance, current deep learning solutions still depend on a carefully annotated dataset for supervised learning. To alleviate time-consuming annotations of body compositions at the pixel level, we used a training-while-annotating strategy. Our experience showed that this strategy could significantly improve the efficiency of data annotations. Based on our experiment in Fig. 3, if the 'training-while-annotating' strategy was not used, it would have taken 70 days to annotate all of these CT scans.

We did not develop novel CNN models or architectures for segmenting body composition in this study. Our experiments showed that there were limited differences among the three classical CNN models in segmentation performance. Despite the reported advantages in relevant publications[36,37], the R2UNet and UNet++ did not demonstrate superior performance versus the classical UNet. Given the "black-box" characteristics of CNN models, the underlying reasons are not clear. A possible explanation is that the batch size (n=4) for training the UNet was larger than the batch size (n=2) for training the R2UNet since all other training parameters were the same. In our implementation, the UNet model had 19,078,402 parameters, the UNet++ model had 27,473,634, and R2Unet had 58,912,098 parameters. This suggests that (1) CNN models with more parameters do not always result in better performance, and (2) batch size is an important parameter for training a CNN model. Hence, we did not pursue developing novel CNN architectures. Based on our results, the CNN models demonstrated promising performance in segmenting the primary body tissues on CT images. Whether this performance is sufficient for relevant clinical applications needs further study. Additionally, we found that the CNN models trained for simultaneous segmentation of the five body compositions demonstrated superior performance versus those trained for segmenting each tissue separately (Tables 2-3). Whether this observation is applicable to other medical image segmentation problems is not clear and also needs additional investigative efforts.

Traditionally, automated segmentation of body compositions is a very challenging problem. First, the densities of various body compositions vary across subjects and CT scans. Although the tissues depicted on CT scans typically have certain densities, this may be associated with the CT acquisition protocols (e.g., dosage, scanner, and body region). It is challenging to use pre-defined thresholds that reliably identify these tissues. Second, the location of body compositions (e.g., VAT) could vary significantly across subjects. It is challenging for the traditional template-based method[27] to accurately and reliably identify these body tissues. In many cases, the boundaries between these tissues are fuzzy, making manual annotation difficult. In recent years, the emergence of CNN-based deep learning (DL) technology makes this segmentation problem much easier to manage without explicitly defining any image features or involving any predefined thresholds. Most of the available methods for segmenting body compositions on CT images were based on either 2-D or 3-D CNN. Dabiri et al.[20] described a 2-D CNN method to automatically segment SM at L3 and T4. Nowak et al.[21] annotated the image slices at L3 and L4 and trained a 2-D CNN model to segment SAT, VAT, and SM. Lee et al.[22] trained a CNN model to identify SM by classifying pixels in the background, muscle, and other tissues. Wang et al.[23] developed two different CNN models, one to identify abdominal CT image slices and the second to segment SAT and VAT. Similarly, Bridge et al.[24] trained a CNN model on L3 slices to identify only muscle, subcutaneous fat, and visceral fat. Weston et al.[25]trained a 2-D UNet model to segment SAT, VAT, muscle, and bone at L3. Although Koitka et al.[28] developed a 3-D convolutional neural network to identify the body compositions on abdominal CT images, they still used a threshold method to differentiate the tissues into muscle and adipose. For comparison purposes, we summarized the available methods for segmenting body compositions in Table 4. Among these studies, Huber et al.[38] described a method to segment body compositions on MRI images, and all other studies focused on CT images. As compared with these available methods, our study has several distinctive characteristics, such as (1) a cohort with a mix of whole-body, chest, and abdominal CT scans, (2) 3-D CNN models, and (3) five different body compositions.

We noticed that most of the available methods reported much higher Dice coefficients than our CNN models (Tables 2–4). Based on our findings, the performance difference should be minimal because all the available CNN-based methods used the UNet model and its variants. There could be several explanations for this. First, most of the described approaches are based on 2-D CNN models, which have much smaller learing/training space when compared with 3-D CNN models. Second, other methods tended to use CT scans with relatively similar protocols, while our cohort had a wide range of image quality (e.g., slice thickness and noise level). Finally, the quantity of various body compositions varies across subjects. For example, in a skinny subject (as shown in Fig. 4), a small error in segmenting SAT could result in a low Dice coefficient or Jaccard index.

**Table 4.** Summary of studies related to automated segmentation of body compositions on volumetric images.

| Publications | Subjects | Method | Body Region | Dice coefficient | | | | |
|---|---|---|---|---|---|---|---|---|
| | | | | SAT | VAT | IMAT | SM | Bone |
| Huber et al. [38] | 20 | 2D UNet | whole body | 0.93 | 0.77 | − | 0.83 | − |
| Magudia et al. [39] | 1499 | 2D UNet | L3 | 0.98 | 0.95 | − | 0.97 | − |
| Lee et al.[29] | 100 | 3D & 2D UNet | whole body | − | 0.95 | − | 0.98 | 0.98 |
| Nowak et al.[21] | 1143 | 2D CNN | L3 and L4 | 0.98 | 0.96 | − | 0.95 | − |
| Koitka et al. [28] | 50 | 3D UNet | abdomen | 0.96 | − | − | 0.93 | 0.94 |
| Zopfs et al. [40] | 86 | 2D UNet | abdomen | 0.95 | − | − | − | − |
| Hemke et al. [41] | 200 | 2D UNet | supra-acetabular | 0.97 | − | 0.91 | 0.95 | 0.92 |
| Paris et al. [42] | 893 | 2D CNN | L3 | 0.99 | 0.98 | 0.9 | 0.98 | − |
| Dabiri et al. [20] | 1004 | 2D CNN | L3 and T4 | − | − | − | 0.98 | − |
| Weston et al. [25] | 2512 | 2D UNet | L3 | 0.98 | 0.94 | − | 0.96 | − |
| Burns et al. [43] | 102 | 2D UNet | Multiple slices | − | − | − | 0.94 | − |
| Bridge et al. [24] | 1129 | 2D UNet | L3 | 0.98 | 0.95 | − | 0.97 | − |
| Lee et al. [22] | 400 | 2D CNN | L3 | − | − | − | 0.93 | − |
| Wang et al. [23] | 40 | 2D CNN | Abdomen | 0.98 | 0.92 | − | | − |
| Popuri et al. [27] | 1004 | template | L3 and T4 | − | − | 0.95 | 0.95 | − |
| Kullberg et al. [44] | 50 | shape analysis | thigh region | 0.99 | 0.97 | − | − | − |

SAT – subcutaneous adipose tissue, VAT – visceral adipose tissue, IMAT – intermuscular adipose tissue, SM – skeletal muscle

## 5. Conclusion

We developed and validated a 3-D CNN-based solution to automatically segment five different body compositions depicted on whole body, chest, and abdominal CT scans. Several CNN models were trained based on 3-D image patches, and their performances were compared. We also developed a training-while-annotating strategy to facilitate the efficient development of the CNN models. Our experiments demonstrate the reliability of the CNN models in identifying these body compositions.

**Disclosure Statement:** The authors have no conflicts of interest to declare.

**Acknowledgments:** RD receives funding from VA Career Development Award (CX001771-01A2).

**Data Availability:** Data is available from the corresponding author (RD) by request via email.